\documentclass[aps,prl,twocolumn,groupedaddress,showpacs]{revtex4}
\usepackage{psfig}

\begin{document}

\title{Bubble Relaxation Dynamics in Double-Stranded DNA}

\author{\sc D. J. Bicout$^{1,2}$ and E. Kats$^{1,3}$} 

\affiliation{$^1$Institut Laue-Langevin, 6 rue Jules Horowitz,
B.P. 156, 38042 Grenoble, France \\                                          
$^{2}$Biomathematics and Epidemiology, ENVL, B.P. 83, 
69280 Marcy l'Etoile, France \\
$^3$ 
L. D. Landau Institute for Theoretical Physics, 
RAS, 117940 GSP-1, Moscow, Russia.\\}

\date{\today}

\begin{abstract} 
The paper deals with the two-state (opening-closing of base pairs) 
model used to describe the fluctuation dynamics of a single bubble 
formation. We present an exact solution for the discrete and finite 
size version of the model that includes end effects and derive 
analytic expressions of the correlation function, survival probability 
and lifetimes for the bubble relaxation dynamics. It is shown that 
the continuous and semi-infinite limit of the model becomes a good 
approximation to exact result when $a^N\ll 1$, where $N$ is bubble 
size and $a$, the ratio of opening to closing rates of base pairs, 
is the control parameter of DNA melting. 
\end{abstract}

 
\maketitle

Upon heating, a double stranded DNA (ds DNA) undergoes a denaturation process 
with the formation of bubbles of increasing size and number and, eventually, 
leading to the separation of the two strands \cite{PS70}. On the other hand, 
many of DNA biological activities require the unzipping of the two strands by 
breaking hydrogen bonds between base pairs. Such open regions of complex DNA,  
enclosing up to $10-30$ broken base pairs, represent a first step of the 
transcription processes and are called the transcription bubbles. Several 
theoretical models have been proposed to describe the phenomenon of bubble 
formation (for a review see e.g., \cite{WB85}). However, the issue remains 
unsettled  with various, and even contradictory, results reported in the 
literature. This is indicative of the complexity of the problem which involves 
number of factors (e.g., base pair sequences, molecular environment, 
counterions, and so on) that can influence the denaturation process in various 
ways (see e.g., \cite{PB89,KM00,SS01}). In addition, as an one or quasi-one 
dimensional system, the ds DNA is expected to be very sensitive to thermal 
fluctuations. Therefore, it seems appropriate in a first step to study 
the fluctuations of local breathing or unzipping of a ds DNA that opens 
up bubbles of a few tens of base pairs.

The characteristic dynamics of these local denaturation zones (bubbles)
in the structure of a ds DNA have been recently probed through fluorescence 
correlation spectroscopy \cite{BK98,ALK03}. This is an essential issue not 
only for physiological processes involving ds DNA but also for providing 
insights on the general nature of fluctuations in such systems. 
From a theoretical modeling perspective, however, we have just begun to 
understand these experimental results. 
In their recent paper \cite{ALK03}, Altan-Bonnet, Libchaber and Krichevsky 
(ALK) have presented a measurement of the dynamics of a single bubble 
formation in ds DNA construct. The authors proposed a simple 
discrete and finite size model for the description of the dynamics of 
bubbles while they used a continuous and semi-infinite version of the 
model to fit their experimental data. In this continuous and semi-infinite 
limit, the survival probability of the bubble reads \cite{ALK03}: 
\begin{eqnarray} 
B_{\infty,c}(t)=\left(1+\frac{x}{2}\right)\, 
{\rm erfc}\left[\frac{\sqrt{x}}{2}\right]
 -\left(\frac{x}{\pi}\right)^{1/2}\,{\rm e}^{-x/4}\:, 
\label{Boftc}  
\end{eqnarray}  
where $x=t/\tau_{\infty,c}$ and the bubble lifetime is, 
\begin{eqnarray} 
\tau_{\infty,c}=\frac{(1+a)}{2k_-(1-a)^2}\:\:;\:\:
a=\frac{k_+}{k_-}={\rm e}^{-\varepsilon/{\rm k_BT}}\:,
\end{eqnarray}  
where $k_+$ and $k_-$ are the opening and closing rates of base-pair, 
respectively, $\varepsilon$ the bubble extension energy and ${\rm k_BT}$ 
the thermal energy. In the same spirit, the dynamics of bubble formation 
have been studied in terms of Fokker-Planck equation \cite{HRM03}.  
In this paper, we go one step forward in providing the exact
solution of the generalized ALK model, taking into account both the 
discreteness of the system and the finite size and including end effects. 
Our motivation in this investigation is to provide  analytic expressions 
for bubble relaxation function, relaxation time, and lifetime. Such  
exact solutions may significantly improve data analyzes and be very 
relevant for any systems with arbitrary $\varepsilon$ and size $N$. 

Following ALK, we denote by $b_n(t)$ the probability density of bubbles 
of size $n$ at time $t$ in the system. Assuming that all conformations
of the ds DNA can be described as two states (closed or open), the 
fluctuations dynamics in the number $n$ of open base-pairs in the bubble 
is described by the master equation,
\begin{eqnarray} 
\left\{\begin{array}{ccl}
\frac{db_0}{dt} & = & k_-b_1-k_1b_0 \\
\frac{db_1}{dt} & = & k_1b_0+k_-b_2-(k_++k_-)b_1 \\
\cdots          &   &  \cdots \\
\frac{db_n}{dt} & = & k_+b_{n-1}+k_-b_{n+1}-(k_++k_-)b_n\: \\
\cdots          &   &  \cdots \\
\frac{db_N}{dt} & = & k_+b_{N-1}+k_2b_{N+1}-(k_++k_-)b_N \\
\frac{db_{N+1}}{dt} & = & k_+b_N-k_2b_{N+1} \, ,
\end{array}\right.
\label{dN} 
\end{eqnarray}
where, in addition to the rates $k_\pm $ in ALK model \cite{ALK03}, we 
have explicitly introduced the opening and the closing rates $k_1$ and $k_2$, 
respectively, for opening the first and closing the last pairs since 
two ends of the DNA helix are sealed.

{\bf\em Stationary Distribution:} 
When $k_1\neq 0$ and $k_2\neq 0$, 
Eq.(\ref{dN}) admits a stationary solution given by,
\begin{eqnarray} 
\frac{b_{\rm st}(n)}{b_{\rm st}(0)}=\left\{\begin{array}{ccl}
k_1a^{n-1}/k_- & ; & 1\leq n\leq N\\
k_1a^{N}/k_2 & ; & n=N+1
\end{array}\right.
\label{bst}
\end{eqnarray} 
where $b_{\rm st}(0)=1/\left[1+(k_1Q/k_-)+(k_1a^{N}/k_2)\right]$ 
with $Q=(1-a^N)/(1-a)$. The equilibrium fraction of DNA molecules that 
are closed, open and with bubbles in the system are given by 
$b_{\rm st}(0)$, $b_{\rm st}(N)$, and $f_b$, respectively, where
\begin{eqnarray}
f_b=\sum_{n=1}^{N}b_{\rm st}(n)=\left(\frac{k_1Q}{k_-}\right)\,
b_{\rm st}(0)\:.
\end{eqnarray}
The equilibrium constants $K_1$ and $K_2$ for the concentrations of species 
in the reactions in Fig.~\ref{fig1} are:
\begin{equation}
K_1=\frac{\mbox{[bubble]}}{\mbox{[closed]}}=\frac{k_1}{k_{\rm b}}\:\:\mbox{and}\:\:
K_2=\frac{\mbox{[open]}}{\mbox{[bubble]}}=\frac{k_{\rm f}}{k_2}\:.
\end{equation}
where the backward $k_b$ and forward $k_f$ rates are, 
\begin{equation}
k_{\rm b}=\frac{k_{\rm f}}{a^N}
=k_-\,\left[\frac{1-a}{1-a^N}\right]\:.
\end{equation}
When $k_1=k_2=0$, the concentration of bubbles tends zero and we have 
$\mbox{[open]}/\mbox{[closed]}=a^N$.

{\bf\em Relaxation Function:} 
To study the fluctuations of bubbles, we consider 
$\Pi(z,t|n_0)=\sum_{n=0}^{N+1}z^n\,b_n(t|n_0)$ 
(where $b_n(t|n_0)$ is conditional the probability density of finding a 
DNA molecule with a bubble of size $n$ at time $t$ given that the size  was 
$n_0$ at time $t=0$) the characteristic function for the system prepared 
with the initial condition, $b_n(t=0|n_0)=\delta_{n,n_0}$. The Laplace 
transform [$\widehat{\Pi}(z,s|n_0)=\int_{0}^{\infty}\!dt\,\Pi(z,t|n_0)\,
{\rm e}^{-st}$] of $\Pi(z,t|n_0)$ is obtained as,
\begin{eqnarray} 
& & \widehat{\Pi}(z,s|n_0)=\frac{1}{D(z,s)}\,\times
\left\{- z^{n_0+1}\right. \nonumber \\
& & +\left[D(z,0)+k_1(1-z)z\right]\,\widehat{b}_0(s|n_0) \nonumber \\
& & +\left.\left[D(z,0)-k_2(1-z)\right]z^{N+1}\,
\widehat{b}_{N+1}(s|n_0)\right\}
\label{Pi}
\end{eqnarray} 
where $D(z,s)=k_+[z-z_1(s)][z-z_2(s)]$ and 
$z_{1,2}(s)=\left[s/k_-+1+a\mp\sqrt{(s/k_-+1+a)^2-4a}\right]/2a$. 
The functions $\widehat{b}_0(s|n_0)$ and $\widehat{b}_{N+1}(s|n_0)$, obtained 
by requiring that the numerator of $\widehat{\Pi}(z,s|n_0)$ cancels at 
the roots of $D(z,s)$, are given by,
\begin{eqnarray}
\widehat{b}_0(s|n_0)=\frac{1}{\Delta}\,\left|
\begin{array}{ccc}
z_1^{n_0} & & \left[sz_1-k_2(1-z_1)\right]z_1^N \\ 
\\
z_2^{n_0} & & \left[sz_2-k_2(1-z_2)\right]z_2^N \\
\end{array}\right| 
\end{eqnarray}
and 
\begin{eqnarray}
\widehat{b}_{N+1}(s|n_0)=\frac{1}{\Delta}\,\left|
\begin{array}{ccc}
s+k_1(1-z_1) & & z_1^{n_0}  \\ 
\\
s+k_1(1-z_2) & & z_2^{n_0}\\
\end{array}\right| 
\end{eqnarray}
with
\begin{eqnarray}
\Delta(s)=\left|
\begin{array}{ccc}
s+k_1(1-z_1) & & \left[sz_1-k_2(1-z_1)\right]z_1^N \\ 
\\
s+k_1(1-z_2) & & \left[sz_2-k_2(1-z_2)\right]z_2^N \\
\end{array}\right|\:. 
\end{eqnarray}

To fit with the experimental conditions by ALK, we assume that the 
system is prepared in the initial conditions 
$b_{\rm st}(n_0)/f_b$ for $1\leq n_0\leq N$ and zero otherwise. 
The quantity of interest is the correlation function $C_N(t)$ that 
describes fluctuations in the bubble population at equilibrium and 
is measured by fluorescence correlation spectroscopy method \cite{ALK03},  
\begin{eqnarray}
& C_N(t) &=\sum_{n_0=1}^{N}\sum_{n=1}^{N}\frac{\left[b_n(t|n_0)-
b_n(\infty|n_0)\right]\,b_{\rm st}(n_0)}{f_b(1-f_b)} \nonumber \\
& & =1-\sum_{n_0=1}^{N}\frac{\left[
b_0(t|n_0)+b_{N+1}(t|n_0)\right]\,b_{\rm st}(n_0)}{f_b(1-f_b)}
\label{coft}
\end{eqnarray}
in which $b_n(0|n_0)=\delta_{n,n_0}$, and we have used the conservation 
of the probability density, $\sum_{n=0}^{N+1}b_n(t|n_0)=1$. Note 
that $C_N(0)=1$ since $b_n(\infty|n_0)=b_{\rm st}(n)$ and $C_N(\infty)=0$. 
Performing the summation in Eq.(\ref{coft}), we find the Laplace transform 
of $C_N(t)$ as, 
\begin{eqnarray}
& & \widehat{C}_N(s)=\frac{1}{s}-\left[\frac{k_-}{(1-f_b)Q}\right]
\,\times 
\nonumber \\
\label{Bofs}
& & \frac{\left[(1-z_1)F(z_2)-(1-z_2)F(z_1)\right]}{s}\:,
\end{eqnarray}
where $F(z)=$ \\ $(1-z^N)\left[s(1-z^{N+1})+(1-z)(k_1+k_2 z^N)\right]
/[z^N\,\Delta]$. From this, the bubble relaxation time is obtained as 
$\tau_N =\widehat{C}(s=0)$. Two limiting cases are considered 
depending on $k_1$ and $k_2$.

{\em $\bullet$  $(k_1+k_2)>0$ limit:} In this case, the bubble 
relaxation time is given by, 
\begin{eqnarray} 
\tau_N & = & \left\{\frac{(1+a^{N+1})}{(1-a)^2}\,\left[
\frac{k_1+k_2}{a^Nk_1k_-+k_2k_-+Qk_1k_2}\right] \right.
\nonumber \\
& - & \frac{2Na^N}{(1-a)(1-a^N)}\,\left[\frac{k_1+k_2}
{a^Nk_1k_-+k_2k_-+Qk_1k_2}\right] \nonumber \\
& + & \left(\frac{1-a^{N+1}}{1-a}\right)\,\left[
\frac{k_-}{a^Nk_1k_-+k_2k_-+Qk_1k_2}\right] 
\nonumber \\
& - & \left.\left(\frac{1-a^{N+1}}{1-a}\right)\,
\left[\frac{1}{a^Nk_1+k_2}\right]\right\}\,{\rm H}(k_1+k_2)\:,
\label{taua}  
\end{eqnarray} 
where $\rm H$ is Heaviside step function defined as $\rm {H}(z)=0$ for 
$z<0$ and $\rm {H}(z)=1$ for $z>0$. When either $k_1$ or $k_2$ tends to 
zero, $\tau_N$ linearly decreases respectively with either $k_1$ or 
$k_2$ towards $\tau_N(0)$ defined as, 
\begin{eqnarray} 
k_-\tau_N(0) & = & \left[\frac{(1+a^{N+1})(1-a^N)-2Na^N (1-a)}
{(1-a)^2(1-a^N)}\right] 
\nonumber \\
& \times & \left\{\begin{array}{ccc}
1 & ; & k_1=0\:,\:k_2>0 \\
a^{-N} & ; & k_1>0\:,\:k_2=0 
\end{array}\right.
\label{tauaz}  
\end{eqnarray} 
Note that $\tau_N(0)$ is independent of $k_1$ and $k_2$ because the 
kinetics in these limits is dominated by the bubbles decay. As 
$N\rightarrow\infty$, the fluctuations of bubbles become independent 
of $N$ with the relaxation function, 
\begin{eqnarray}
\label{Bofs1}
\widehat{C}(s)=\frac{1}{s}- \frac{k_-(1-a)(1-z_1)}
{(1-f_b)s\left[s+k_1(1-z_1)\right]}\:,
\end{eqnarray} 
and lifetime,
\begin{eqnarray} 
\tau_{\infty}=\frac{1}{(1-a)\left[(1-a)k_-+k_1\right]}\:.
\label{tauinf}  
\end{eqnarray}  

\begin{figure}[ht] 
\vspace{0.3cm}
\centerline{\vbox{
\psfig{figure=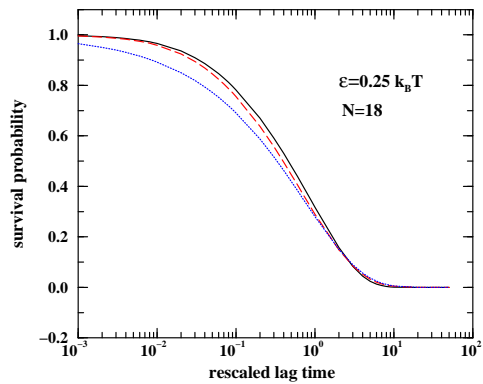,width=2.5in,angle=0}
}}
\vspace{-0.25cm}
\caption{Bubble survival probabilities, from the top to the bottom, 
$B_N(t)$ (solid line), $B_{\infty}(t)$ (long-dashed line) and 
$B_{\infty,c}(t)$ (dotted line) versus the rescaled lag times 
$t/\tau_N$, $t/\tau_{\infty}$ and $t/\tau_{\infty,c}$, respectively.}
\label{fig2}
\end{figure}

\begin{figure}[ht]
\vspace{0.3cm}
\centerline{\vbox{
\psfig{figure=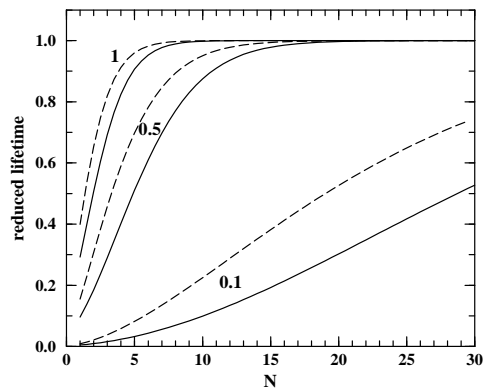,width=2.5in,angle=0}
}}
\vspace{-0.25cm}
\caption{Reduced lifetime, $\tau_N/\tau_{\infty}$ in Eq.(\ref{taua}) 
for $k_1=0$ (dashed line) and Eq.(\ref{taub}) (solid line), 
as a function of bubble size, $N$. Quoted numbers represent the bubble 
extension energy $\varepsilon/{\rm k_BT}$.}
\label{fig3}
\end{figure}

{\em $\bullet$ $k_1=k_2=0$ limit:}
In this case, $\widehat{C}_N(s)=\widehat{B}_N(s)$, where $B_N(t)$ is 
the survival probability of bubbles. Likewise, the bubble 
lifetime $\tau_N=\widehat{B}_N(s=0)$ is given by,
\begin{eqnarray} 
\tau_N=\frac{(1-a^N)(1-a^{N+2})-N(N+2)(1-a)^2a^N}
{k_-(1-a)^2(1-a^N)(1-a^{N+1})}\:. 
\label{taub}  
\end{eqnarray}  
When  $N\rightarrow\infty$, Eq.(\ref{Bofs}) reduces to 
$\widehat{B}_{\infty}(s)=(1/s)-k_-(1-a)\,[1-z_1]/s^2$, and, 
\begin{eqnarray} 
& & B_{\infty}(t)=1-\frac{x}{1-a}+  \nonumber \\ 
& & \frac{(1-a)}{2a}\,\int_{0}^{y}\!dz\,\left( \frac{y}{z}-1\right)\, 
\exp\left[-\frac{(1+a)}{2\sqrt{a}}\,z\right]\, 
{\rm I}_1[z]\:, 
\label{Boftb}  
\end{eqnarray}  
where ${\rm I}_1[\cdots]$ is the modified Bessel function of order one, 
$y=2x\sqrt{a}/(1-a)^2$ and $x=t/\tau_{\infty}$. It is worth noting 
that even in the $N \to \infty $ limit the 
exact solution Eq.(\ref{Boftb}) for the bubble survival
probability is different from Eq.(\ref{Boftc}) given in
\cite{ALK03}. The fact is that, depending on the size $N$ and the 
parameter ``$a$'', the discreteness of the system is an ingredient which 
might be taken into account to capture the correct bubble dynamics. 
This is illustrated in Fig.~\ref{fig2} where the exact survival 
probability is compared with its $N \to \infty $ limit and the ALK 
continuous model. Figure~\ref{fig3} shows the departure in the bubble 
lifetime to the continuous limit as a function of bubble size. It 
clearly appears from Figs.~\ref{fig2} and \ref{fig3} that the 
continuous limit as done by ALK \cite{ALK03} becomes a fairly 
good approximation to exact result for $a^N\ll 1$ (where $a \leq 1$ 
is the control parameter for the ds DNA melting  \cite{ALK03,HM03}).
 
Simple inspection of expressions in Eqs.(\ref{Bofs}), (\ref{taua}) and 
(\ref{taub}), and of the figures, indicate that the behavior of bubble 
dynamics is controlled by the bubble size $N$ and the parameter $a$  
(ratio of opening to closing rates of base-pairs). As $a\leq 1$ according 
to the experimental situation in \cite{ALK03}, the closing of bubbles 
is the fastest process in the bubbles kinetics. The parameter $a$ also 
controls the denaturation transition. As $a \to 1$, there is a kind of 
''critical slowing down'' where the fluctuations of bubbles are 
described by an unbiased diffusion process. For instance, the bubble lifetime 
in Eq.(\ref{taub}) reduces to,
\begin{equation}
\tau_N=\frac{(N+1)(N+2)}{12k_-}\:,
\end{equation}
in the $a \to 1$ limit, and $\tau_N$ diverges with the bubble size. 

It may be useful for practical purposes to have an idea of numerical 
values of physical parameters entering in the problem. In the absence of 
direct measurement of $k_-$, for instance, one can use the experimental 
data in \cite{ALK03} in conjunction with theoretical results to 
estimate the closing rate $k_-$. The results of such an estimation 
are presented in Table~\ref{tb1}.

\begin{table} 
\caption{Estimate of $k_-$ using the expressions of the bubble lifetime 
in the case of $k_1=k_2=0$. 
In Ref.~\cite{ALK03}, the experimental bubble lifetime is equal to 
$95\,\mu s$ at T=303K for $N=18$ and  DNA samples $M_{18}$ and $A_{18}$.}
\begin{tabular}{l|c|ccccc}  \hline\hline
\multicolumn{2}{l|}{$\varepsilon/{\rm k_BT}$} & $0.1$   & & $0.5$ & & $1$  \\ \hline
\multicolumn{2}{c|}{Lifetime ($\mu s$)} & \multicolumn{5}{c}{$k_-\:\:(\times 10^6\,{\rm s}^{-1})$}\\ \hline
$\tau_N$ & $95$          & $0.300$ & & $0.0675$ & & $0.0263$  \\
$\tau_{\infty}$ & $95$   & $1.162$ & & $0.0680$ & & $0.0263$  \\
$\tau_{\infty,c}$ & $95$ & $1.110$ & & $0.0550$ & & $0.0180$  \\ \hline \hline
\end{tabular} 
\vspace{0.5cm}
\label{tb1} 
\end{table}

To summarize, we have presented an exact solution of the discrete and 
finite size model in Eq.(\ref{dN}) for the description of the fluctuations 
dynamics of bubble formation. The twofold merit of this two-state 
(open and closed) model is to already include sufficient complexity of the 
bubble dynamics over biomolecular relevant scales and to allow exact 
analytical solution. The mains results of the paper are the 
expressions in Eqs.(\ref{Bofs}), (\ref{taua}) and (\ref{taub}) for the 
bubble correlation function, relaxation time and bubble lifetime, 
respectively. These results, consistent with available data, may prove 
to be useful for analysis and interpretation of experimental data 
on bubble fluctuations and they are amenable for further experimental 
tests. It is worthwhile to mention in addition that different expressions 
for the relaxation function and time can be generated within the 
theoretical framework developed above by simply using different 
initial conditions in Eq.(\ref{coft}) for the preparation of the system. 

Given the closing and opening rates of base pair, the model discussed 
above allows also to study phenomena related to the denaturation 
mechanisms of DNA such as heating, changing buffer surrounding, or 
applying external torques or forces \cite{MS95,CM99,LN00,CK02}. Likewise, 
the model can easily modified to include more than two states in order to 
describe, for instance, the intermediates states between bond and broken 
states. Finally, although the calculations may become more involve 
and intricate, the theory outlined above can be extended in several 
directions in including in Eq.(\ref{dN}), for example, the effects of 
base pair sequence in the opening and closing rates (two and three 
hydrogen bonds being involved in A-T and G-C base pairs, respectively), 
initiation of several bubbles, bubbles fission and fusion processes, 
and so on.

\acknowledgements

E.K. acknowledges partial support of this work by 
\newline
INTAS grant 01-0105.


\begin{references} 

\bibitem{PS70} D. Poland, H. A. Sheraga, Theory of Helix - Coil Transitions
in Biopolymers, Academic Press, New York (1970).

\bibitem{WB85} R. M. Wartell, A. S. Benight, Phys. Rep., {\bf 126}, 67 (1985).

\bibitem{PB89} M. Peyrard, A. R. Bishop, Phys. Rev. Lett., {\bf 62}, 2755
(1989).

\bibitem{KM00} Y. Kafri, D. Mukamel, L. Peliti, Phys. Rev. Lett.,
{\bf 85}, 4988 (2000).

\bibitem{SS01} N. Singh, Y. Singh, Phys. Rev. E, {\bf 64}, 042901 (2001). 

\bibitem{BK98} G. Bonnet, O. Krichevskii, A. Libchaber, 
Proc. Natl. Acad. Sci. USA, {\bf 95}, 8602 (1998).

\bibitem{ALK03} G. Altan-Bonnet, A. Libchaber and O. Krichevsky, 
Phys. Rev. Lett., {\bf 90}, 138101-1 (2003).

\bibitem{HRM03} A. Hanke and R. Metzler, 
J. Phys. A: Math. Gen. {\bf 36}, L473 (2003).

\bibitem{HM03} T. Hwa, E. Mariani, K. Sneppen, L-h. Tang, 
Proc. Natl. Acad. Sci. USA, {\bf 100}, 4411 (2003).

\bibitem{MS95} J. F. Marko, E. D. Siggia, Phys. Rev. E, {\bf 52}, 2912 (1995);
R. M. Fye, C. J. Benham, Phys. Rev. E. {\bf 59}, 3408 (1999).

\bibitem{CM99} S. Cocco, R. Monasson, Phys. Rev. Lett., {\bf 83}, 5178 (1999).

\bibitem{LN00} D. K. Lubensky, D. R. Nelson, 
Phys. Rev. Lett., {\bf 85}, 1572 (2000).

\bibitem{CK02} J. Chuang, Y. Kantor, M. Kardar, 
Phys. Rev. E, {\bf 65}, 011802 (2002).

\end{references}
\end{document}